\documentclass{article}
\usepackage{spconf,amsmath,graphicx,hyperref}
\usepackage{booktabs} 
\usepackage{amssymb}
\usepackage{pifont}
\usepackage{multirow}
\usepackage{url}

\title{ALIGNING LANGUAGE MODELS FOR LYRIC-TO-MELODY GENERATION WITH RULE-BASED MUSICAL CONSTRAINTS}
\name{Hao Meng$^1$ \qquad Siyuan Zheng$^1$ \qquad Shuran Zhou$^1$ \qquad Qiangqiang Wang$^1$ \qquad Yang Song$^1$} 
\address{$^1$Zuoyebang Education Technology, Beijing, China\\
\{menghao05, zhengsiyuan02, zhoushuran, wangqiangqiang, songyang\}@zuoyebang.com}
\begin{document}
\ninept
\maketitle
\begin{abstract}
Large Language Models (LLMs) show promise in lyric-to-melody generation, but models trained with Supervised Fine-Tuning (SFT) often produce musically implausible melodies with issues like poor rhythm and unsuitable vocal ranges, a phenomenon we term "constraint violation". To address this, we propose a novel alignment framework that instills musical knowledge without human annotation. We define rule-based musical constraints to automatically generate a preference dataset from an SFT model's outputs. The model is then aligned through a sequential process, first using Direct Preference Optimization (DPO) on paired preference data, followed by Kahneman-Tversky Optimization (KTO) on unpaired negative samples. Experimental results demonstrate that our aligned model substantially reduces rule violations and outperforms strong baselines in both objective and subjective evaluations, generating melodies with substantially improved musicality and coherence. An interactive demo with audio comparisons is available at \url{https://arain233.github.io/AligningMelody-demo}.
\end{abstract}

%

\section{Introduction}
\label{sec:intro}

The advent of Large Language Models (LLMs) has catalyzed a paradigm shift across numerous domains of artificial intelligence, from natural language understanding to complex reasoning \cite{brown2020language}. In the realm of creative arts, LLMs are increasingly being leveraged for generative tasks, spanning from large-scale raw audio generation \cite{dhariwal2020jukebox} to the composition of symbolic music \cite{ding2024songcomposer, yu2024songglm}. Lyric-to-melody generation, a core challenge in automatic songwriting, has emerged as a particularly promising application area, offering significant value from providing creative inspiration to both amateur enthusiasts and professional musicians. 

Moreover, with the rise of voice-based interaction, singing is becoming a crucial component for more expressive voice agents. Consequently, leading conversational AIs like Step-Audio \cite{huang2025step}, GLM-4-Voice \cite{zeng2024glm}, and Doubao \cite{bai2024seed}, have incorporated this capability. However, while these models can vocalize simple phrases, they often produce musically implausible or aesthetically unappealing melodies when faced with creative or complex lyrics. This highlights that high-fidelity generation requires a nuanced understanding of both linguistic semantics and musical structure \cite{ji2023survey}.

Early approaches to this task often relied on encoder-decoder architectures, such as SongMASS \cite{sheng2021songmass}, or template-based systems like TeleMelody \cite{ju2022telemelody, zhang2022relyme}, which aimed to bridge the gap between lyrics and melody through intermediate representations. More recently, the powerful sequence modeling capabilities of LLMs have been directly applied to this problem. Models like SongComposer \cite{ding2024songcomposer} and SongGLM \cite{yu2024songglm} have demonstrated that by fine-tuning a pretrained LLM on lyric-melody pairs, it can learn to generate coherent melodies in an end-to-end, autoregressive manner.

Despite these advances, the standard Supervised Fine-Tuning (SFT) paradigm has a critical limitation: it learns to imitate the statistical patterns in the training data but lacks robust adherence to musical principles. Consequently, SFT-based models often generate melodies with musically implausible artifacts, or constraint violations \cite{ji2023survey,ding2024songcomposer}. These include monotonous pitch sequences, rhythmically awkward note durations, melodies that fall outside a comfortable human vocal range, or failures to correctly align lyrics with notes. Such flaws render the generated music unsuitable for real-world deployment, such as providing inspiration for musicians or powering interactive voice agents.

Therefore, to move beyond simple imitation and correct these flaws, it is essential to introduce a subsequent alignment stage that refines the model's output based on human or expert-defined musical preferences  \cite{ziegler2019fine}. To steer model behavior towards human preferences, Reinforcement Learning from Human Feedback (RLHF) \cite{christiano2017deep} has become a standard alignment technique. However, RLHF is notoriously complex and computationally expensive, and its reliance on large-scale human annotation creates a significant bottleneck \cite{ouyang2022training}. This work posits that for domains where basic symbolic-level principles--such as constraints on pitch range, note duration, and vocal register--are well-defined, expert knowledge can be codified into deterministic rules, providing a scalable and cost-effective alternative to human feedback.

We introduce a novel framework for aligning LLMs for lyric-to-melody generation using rule-based musical constraints. Our main contributions are:
\begin{itemize}
    \item A comprehensive set of five rule-based musical constraints that formalize fundamental principles of melody writing, targeting common failure modes of generative models.
    \item A fully automated pipeline for generating a large-scale preference dataset. By using our rules to evaluate the SFT model's outputs, we create both paired (winner, loser) and unpaired (loser-only) data without any human intervention.
    \item A sequential alignment strategy that first applies Direct Preference Optimization (DPO) and then Kahneman-Tversky Optimization (KTO) to effectively learn from the entire spectrum of preference signals.
\end{itemize}

Extensive experiments demonstrating that our method substantially reduces musical errors and achieves state-of-the-art performance on both objective and subjective metrics, validating the efficacy of rule-based alignment as a scalable and cost-effective paradigm for instilling domain-specific knowledge into generative models.

\begin{figure*}[t]
  \centering
  \includegraphics[width=\textwidth]{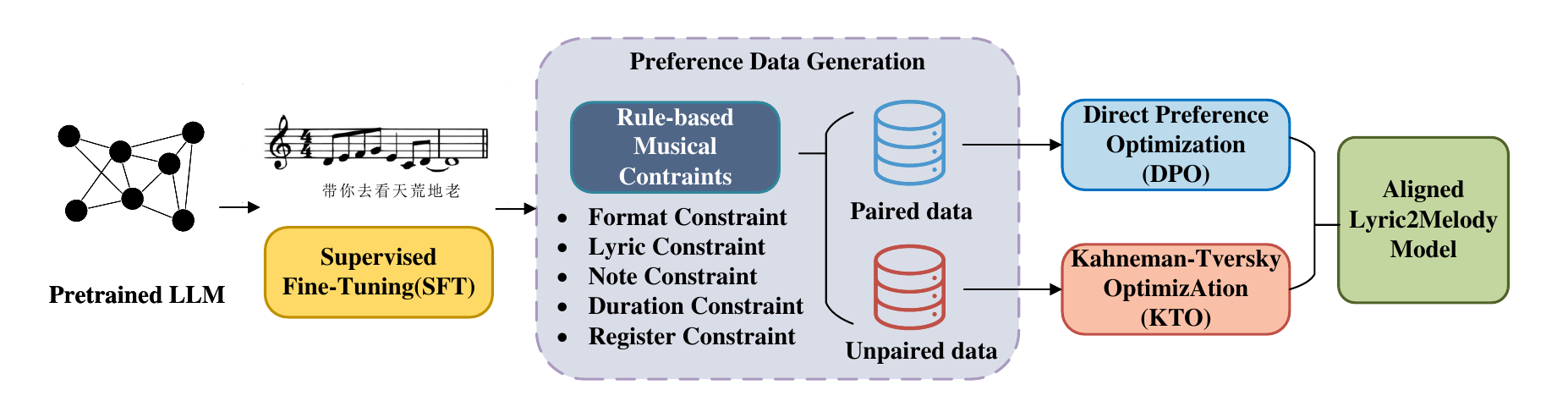}
  \caption{The overall framework of our proposed method. An LLM is first fine-tuned via SFT. Then, a preference dataset is automatically generated by evaluating the SFT model's outputs against a set of rule-based musical constraints. Finally, the model is aligned using a sequential DPO and KTO process.}
  \label{fig:framework}
\end{figure*}

\section{Method}
\label{sec:method}

\subsection{Framework Overview}
\label{ssec:overview}

Our proposed framework aligns a pretrained LLM for high-quality lyric-to-melody generation through a three-stage process, as illustrated in Figure \ref{fig:framework}.

\textbf{1. Supervised Fine-Tuning (SFT):} We begin with a pretrained LLM and fine-tune it on a large corpus of paired lyric-melody data. This initial stage equips the model with the fundamental ability to map lyrical input to a melodic output in the specified symbolic format.

\textbf{2. Preference Data Generation:} The SFT model is then used to generate multiple melody candidates for a large, diverse set of unseen lyrics. Each generated melody is evaluated against our set of rule-based musical constraints. Based on this evaluation, we automatically construct a preference dataset containing both paired data (a rule-compliant "winner" and a rule-violating "loser") and unpaired data (collections of rule-violating outputs for prompts where no compliant melody was generated).

\textbf{3. Sequential Alignment:} Finally, we perform a post-training alignment phase on the SFT model. We employ a sequential optimization strategy that first refines the model with DPO \cite{rafailov2023direct} on the paired data and then further tunes it with KTO \cite{ethayarajh2024kto} on the unpaired negative samples. This process fine-tunes the model to prefer musically plausible outputs, resulting in our final aligned Lyric2Melody model.

\subsection{Symbolic Melody Representation}
\label{ssec:representation}
We adopt a human-readable and machine-parsable symbolic format to represent melodies, drawing inspiration from previous works \cite{ding2024songcomposer}. A melody is tokenized into a sequence of note events delimited by $  |  $. Each note event N is formally defined as a tuple $  N = (l, p, d)  $, where $  l  $  represents the lyric syllable (or - for a melisma), $ p $ is the MIDI pitch number, and $ d $ is the note duration in milliseconds. This structured representation ensures a precise alignment between the lyrical content and the melodic contour, providing a clear and effective format for the language model to process and generate.

\subsection{Rule-based Musical Constraints}
\label{ssec:rules}
To formalize musical common sense, we define five categories of constraints that target frequent and perceptually jarring errors in generated melodies.

\textbf{1. Format Constraint:} This is a fundamental syntactic check to ensure the model's output adheres to the defined symbolic representation. The output must be correctly parsable into a sequence of `(lyric, pitch, duration)` tuples.

\textbf{2. Lyric Constraint:} The generated melody must accurately correspond to the input lyrics. Let $L_{in} = (w_1, w_2, \dots, w_m)$ be the sequence of words in the input lyric. Let $L_{out}$ be the sequence of non-melisma lyric tokens extracted from the generated output. This constraint requires that $L_{out}$ is a valid segmentation of $L_{in}$.

\textbf{3. Note Constraint (Monotony Avoidance):} To prevent musically uninteresting melodies dominated by a single pitch, we constrain the amount of consecutive note repetition. Let $P = (p_1, p_2, \dots, p_n)$ be the sequence of pitches in the generated melody. The constraint is satisfied if the ratio of consecutive identical pitches does not exceed a threshold $\tau_{note}$:
\begin{equation}
\frac{\sum_{i=1}^{n-1} \mathbb{I}(p_i = p_{i+1})}{n-1} \le \tau_{note},
\end{equation}
where $\mathbb{I}(\cdot)$ is the indicator function.

\textbf{4. Duration Constraint (Rhythmic Plausibility):} This rule ensures that note durations are rhythmically sensible and performable. It comprises two conditions:
\begin{itemize}
    \item \textbf{Note Length:} Each note duration $d_i$ must fall within a perceptually valid range: $d_{min} \le d_i \le d_{max}$. This prevents notes from being too short to be heard or unnaturally long.
    \item \textbf{Final Note Length:} The final note of a musical phrase typically has a longer, more conclusive duration. We enforce a separate, longer range for the final note's duration, $d_{n}$.
\end{itemize}

\textbf{5. Register Constraint (Vocal Range):} To ensure the generated melody is singable by an average person, all pitches must lie within a typical human vocal range. Let $P$ be the pitch sequence. The constraint requires:
\begin{equation}
p_{min} \le p_i \le p_{max}, \quad \forall p_i \in P,
\end{equation}
where $[p_{min}, p_{max}]$ is a predefined MIDI note range (e.g., C4 to C6).

\begin{table*}[t]
\centering
\caption{Objective evaluation of different Lyric-to-Melody generation methods on English and Chinese test sets.}
\label{tab:main_comparison}
\begin{tabular*}{\linewidth}{@{\extracolsep{\fill}}l|ccc|ccc} 
\toprule
\textbf{Method} & \multicolumn{3}{c|}{\textbf{English}} & \multicolumn{3}{c}{\textbf{Chinese}} \\
\cmidrule{2-7}
& \textbf{PD(\%) $\uparrow$} & \textbf{DD(\%) $\uparrow$} & \textbf{MD $\downarrow$} & \textbf{PD(\%) $\uparrow$} & \textbf{DD(\%) $\uparrow$} & \textbf{MD $\downarrow$} \\
\midrule
SongMASS & 30.11 & 19.61 & \textbf{1.87} & - & - & - \\
TeleMelody & 30.08 & 31.51 & 3.41 & 25.08 & 35.09 & 3.25 \\
TeleMelody(RelyMe) & 31.27 & 30.99 & 3.32 & 27.59 & 34.70 & 3.29 \\
SongComposer & 31.58 & 31.44 & 3.31 & 30.79 & 33.68 & 3.11 \\
\midrule
\textbf{Proposed} & \textbf{32.37} & \textbf{37.11} & 2.63 & \textbf{33.94} & \textbf{43.44} & \textbf{2.58} \\
\bottomrule
\end{tabular*} 
\end{table*}

\subsection{Sequential Alignment with DPO and KTO}
\label{ssec:alignment}
Our alignment strategy is designed to maximize the learning signal extracted from the automatically generated preference data. For a given input lyric $x$, we generate $k$ candidate melodies. If this set contains at least one rule-compliant melody $y_w$ (winner) and at least one rule-violating melody $y_l$ (loser), we form a preference pair $(x, y_w, y_l)$ for DPO training. However, for some prompts, the SFT model may fail to generate any rule-compliant melodies. A DPO-only approach would discard these instances, wasting valuable data about the model's failure modes. To address this, we collect these rule-violating outputs as an unpaired dataset of "undesirable" responses $\{y_u\}$ and use KTO to learn from them in a subsequent training stage. This sequential approach ensures data efficiency and robustly targets the model's weaknesses.

\textbf{Direct Preference Optimization (DPO):} DPO directly optimizes the policy model $\pi_\theta$ to satisfy preferences by maximizing the likelihood of preferred responses over dispreferred ones, without an explicit reward model \cite{rafailov2023direct}. The loss function is defined as:
\begin{multline}
\mathcal{L}_{\text{DPO}}(\pi_\theta; \pi_{\text{ref}}) = -\mathbb{E}_{(x, y_w, y_l) \sim \mathcal{D}_{\text{paired}}} \bigg[ \log \sigma \bigg( \beta \log \frac{\pi_\theta(y_w|x)}{\pi_{\text{ref}}(y_w|x)} \\
 - \beta \log \frac{\pi_\theta(y_l|x)}{\pi_{\text{ref}}(y_l|x)} \bigg) \bigg],
\end{multline}
where $\pi_{\text{ref}}$ is a frozen copy of the initial SFT model, $\beta$ is a hyperparameter controlling the deviation from the reference policy, and $\sigma$ is the logistic function.

\textbf{Kahneman-Tversky Optimization (KTO):} KTO is an alignment method that learns from binary labels of "desirable" or "undesirable" generations, rather than explicit pairs \cite{ethayarajh2024kto}. Since our unpaired dataset $\mathcal{D}_{\text{unpaired}}$ consists solely of undesirable samples, we use the corresponding part of the KTO loss function to discourage the model from producing such outputs:
\begin{multline}
\mathcal{L}_{\text{KTO}}(\pi_\theta; \pi_{\text{ref}}) = \mathbb{E}_{(x, y_u) \sim \mathcal{D}_{\text{unpaired}}} \bigg[ \log \Big( 1 \\
- \sigma \Big( \beta \log \frac{\pi_\theta(y_u|x)}{\pi_{\text{ref}}(y_u|x)} \Big) \Big) \bigg],
\end{multline}
where $y_u$ is an undesirable (rule-violating) response.

Our alignment process is iterative. We first fine-tune the SFT model using the DPO loss on the paired dataset $\mathcal{D}_{\text{paired}}$. Subsequently, the resulting model is further trained using the KTO loss on the unpaired dataset $\mathcal{D}_{\text{unpaired}}$. This sequential approach first refines the model's preferences with high-quality paired data and then robustly discourages common failure modes using the broader set of unpaired negative examples.

\section{Experiment}
\label{sec:experiments}

\subsection{Experimental Setup}
\label{ssec:setup}

\textbf{Datasets:} Our training data for the SFT stage consists of approximately 800k Chinese and 500k English sentence-level lyric-melody pairs, aggregated from the SongComposer dataset and proprietary sources. For evaluation, we curated a test set of 1000 sentences (500 Chinese, 500 English) from the GTSinger dataset, ensuring no overlap with the training set. The preference dataset for alignment was generated by prompting the SFT model with 20k unseen lyrics in both languages. The resulting dataset consisted of approximately 90\% paired data for DPO and 10\% unpaired data for KTO.

\begin{table}[t]
\centering
\caption{Mean Opinion Score (MOS) evaluation for overall musical quality. GT refers to ground truth recordings.}
\label{tab:mos_evaluation}
\begin{tabular}{lc}
\toprule
\textbf{Method} & \textbf{MOS $\uparrow$} \\
\midrule
GT & 3.50 \\
\midrule
SongMASS & 3.18 \\
TeleMelody & 3.09 \\
TeleMelody(RelyMe) & 3.26 \\
SongComposer & 2.92 \\
Step-Audio-TTS & 3.19 \\
\midrule
\textbf{Proposed} & \textbf{3.42} \\
\bottomrule
\end{tabular}
\end{table}

\noindent \textbf{Evaluation Metrics:} We evaluate the models using both objective and subjective metrics.
\begin{itemize}
    \item \textbf{Objective Metrics:} Following previous work \cite{ju2022telemelody}, we use three symbolic metrics that compare generated melodies to ground-truth references: Pitch Distribution Similarity (PD, \%$\uparrow$), Duration Distribution Similarity (DD, \%$\uparrow$), and Melody Distance (MD, $\downarrow$). PD and DD measure the cosine similarity of pitch and duration histograms, respectively, while MD uses Dynamic Time Warping (DTW) to compute the distance between pitch contours. We process the melodic metrics with relative normalization to enhance the fairness of the evaluation.
    \item \textbf{Subjective Metrics:} We conducted a Mean Opinion Score (MOS) test. Melodies were synthesized into singing voice audio using a custom-trained vocoder. Ten volunteers with musical backgrounds rated the sampled 20 audio clips on a scale of 1 to 5 based on overall musical quality, including melodiousness and rhythmic appeal.
\end{itemize}

\begin{table*}[t]
\centering
\caption{Ablation study of the alignment components on English and Chinese test sets.}
\label{tab:ablation_studies}
\begin{tabular*}{\linewidth}{@{\extracolsep{\fill}}l|ccc|ccc} 
\toprule
\textbf{Method} & \multicolumn{3}{c|}{\textbf{English}} & \multicolumn{3}{c}{\textbf{Chinese}} \\
\cmidrule{2-7}
& \textbf{PD(\%) $\uparrow$} & \textbf{DD(\%) $\uparrow$} & \textbf{MD $\downarrow$} & \textbf{PD(\%) $\uparrow$} & \textbf{DD(\%) $\uparrow$} & \textbf{MD $\downarrow$} \\
\midrule
\textbf{Proposed} & \textbf{32.37} & 37.11 & \textbf{2.63} & \textbf{33.94} & \textbf{43.44} & \textbf{2.58} \\
\midrule
DPO & 31.22 & 37.25 & 2.77 & 30.83 & 40.98 & 2.87 \\
KTO & 31.62 & \textbf{37.96} & 2.77 & 28.64 & 40.53 & 3.10 \\
SFT & 30.42 & 36.46 & 2.95 & 27.00 & 40.02 & 3.12 \\
\bottomrule
\end{tabular*}
\end{table*}

\noindent \textbf{Baselines:} We compare our method against several state-of-the-art lyric-to-melody generation systems: SongMASS \cite{sheng2021songmass}, TeleMelody \cite{ju2022telemelody}, TeleMelody optimized based on RelyMe \cite{zhang2022relyme}, and SongComposer \cite{ding2024songcomposer}. For the MOS evaluation, we also include the end-to-end singing voice generation model Step-Audio-TTS \cite{huang2025step} as a strong audio-domain baseline.

\noindent \textbf{Implementation Details:} Our model is based on the Qwen2.5-0.5B pretrained LLM \cite{yang2024qwen2}. The SFT stage was run for 500,000 steps. For the alignment stage, we set the DPO/KTO scaling factor $\beta$ to 0.1 and used the Adam optimizer with a learning rate of $1 \times 10^{-6}$. All models were trained on 8 NVIDIA A800 GPUs. The vocoder for audio synthesis was trained on an internal high-quality singing dataset based on the TechSinger\cite{guo2025techsinger} architecture.

\subsection{Main Results}
\label{ssec:results}

Table \ref{tab:main_comparison} presents the objective evaluation results. Our proposed method consistently outperforms all baselines on PD and DD for both English and Chinese, indicating that the generated melodies more closely match the pitch and rhythm characteristics of human-composed music. It is worth noting that while SongMASS achieves the lowest MD on the English set, its significantly lower DD score suggests this may be an artifact of the DTW alignment algorithm rather than a reflection of superior rhythmic quality. Our model, in contrast, demonstrates a more balanced and robust performance across all metrics, highlighting its ability to generate melodies that are holistically closer to human compositions. This underscores the importance of the subjective MOS evaluation, which captures the integrated perceptual quality of the music.

The results of the subjective MOS evaluation are shown in Table \ref{tab:mos_evaluation}. This evaluation, which directly assesses perceptual quality, provides the most compelling evidence of our method's effectiveness. Our proposed model achieves a MOS of 3.42, significantly surpassing all baseline methods. Notably, this score is very close to the ground truth (GT) audio, which received a score of 3.50, indicating that the melodies generated by our aligned model are perceived by human experts as being of nearly human-level quality.

\subsection{Ablation Studies and Analysis}
\label{ssec:ablation}

To dissect the contributions of our sequential alignment strategy, we conducted an ablation study, the results of which are presented in Table \ref{tab:ablation_studies}. The full proposed method (SFT+DPO+KTO) achieves the best overall performance, confirming the synergistic benefit of our sequential alignment strategy. Both DPO-only and KTO-only alignment provide substantial improvements over the SFT baseline, demonstrating that both components are effective at enhancing musical quality. Interestingly, KTO alone yields the highest DD score, suggesting it is particularly effective at penalizing the rhythmically implausible durations that constitute a large portion of the unpaired negative data. However, the combination in our proposed model yields the best results, particularly in PD and MD, confirming the value of first refining preferences with DPO before robustly pruning failure modes with KTO.

\begin{figure}[t]
  \centering
  \includegraphics[width=\linewidth]{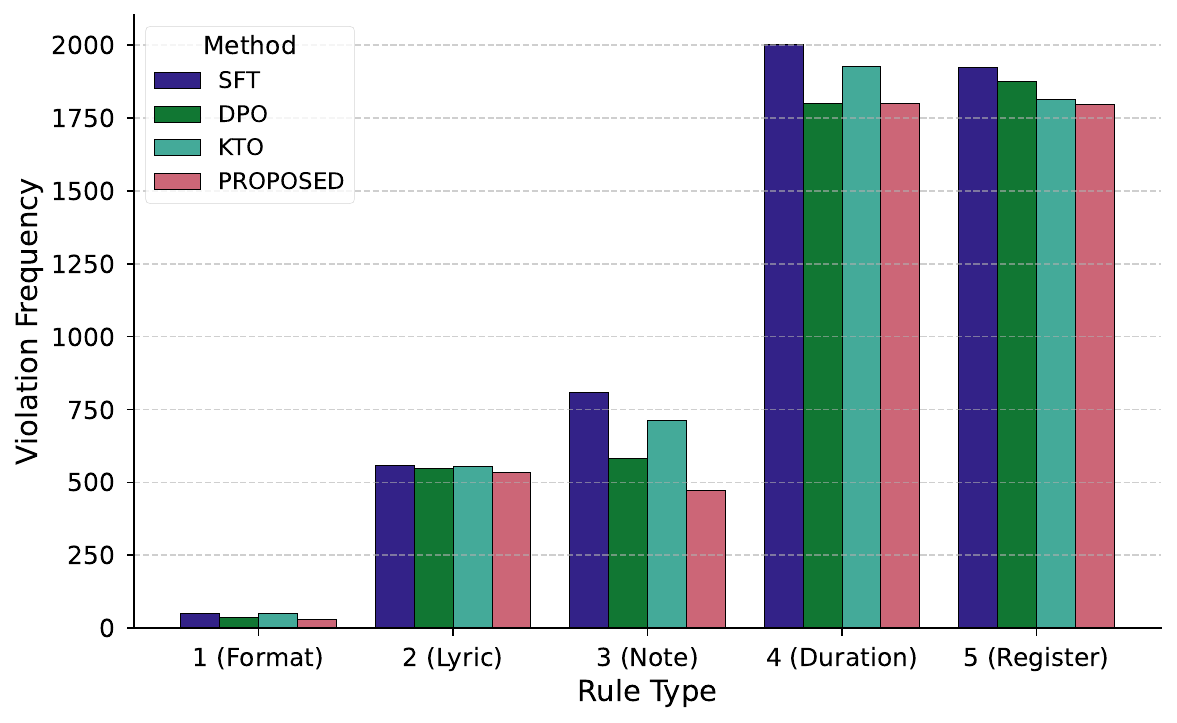}
  \caption{Frequency of rule violations on a held-out test set. Our proposed method substantially reduces violations across all rule types compared to the SFT baseline and ablated models.}
  \label{fig:rule_violations}
\end{figure}

To directly verify that our alignment process successfully teaches the model to follow the specified musical rules, we analyzed the frequency of rule violations on a held-out set of lyrics. Figure \ref{fig:rule_violations} shows the violation counts for each rule category across different models. The SFT baseline commits a large number of errors, particularly for Duration and Register constraints. This is expected, as these rules define wide valid ranges (e.g., typical vocal range or acceptable note lengths), making it statistically probable for unconstrained generations to fall outside these boundaries. Both DPO and KTO individually reduce these violations, but our full proposed method achieves the most dramatic reduction across all five categories. This provides direct evidence that the improvements in objective and subjective scores are driven by the model's learned adherence to fundamental musical principles.

\section{Conclusion}
\label{sec:conclusion}

In this paper, we addressed the critical challenge of musical plausibility in LLM-based lyric-to-melody generation. We introduced a novel alignment framework that uses codified musical constraints to auto-generate preference data for a sequential DPO-KTO process. Our approach  instills musical domain knowledge into the LLM, substantially reducing compositional errors and achieving state-of-the-art results. Evaluations show the generated melodies are superior in  symbolic metrics and human expert perception. This work validates the potential of rule-based alignment as a scalable and effective alternative to human feedback for developing specialized, high-quality models. Future work could explore expanding the rule set to encompass more complex harmonic and structural principles or enabling interactive rule definition for user-controllable melody generation.

\vfill\pagebreak

\bibliographystyle{IEEEbib}
\bibliography{refs}

\end{document}